\documentclass[aps,prb,reprint,showpacs,floatfix]{revtex4-1}
\usepackage{amsmath,graphics,epsfig,color,verbatim}

\let\oldhat\hat

\renewcommand{\hat}[1]{\oldhat{\mathbf{#1}}}

\begin{document}

\title{Electronic correlations in monolayer VS$_2$}
\author{Eric B. Isaacs}
\email{eric.isaacs@columbia.edu}
\author{Chris A. Marianetti}
\email{chris.marianetti@columbia.edu}
\affiliation{Department of Applied Physics and Applied Mathematics, Columbia University, New York, New York 10027, USA}

\begin{abstract}
The layered transition metal dichalcogenide vanadium disulfide
(VS$_2$), which nominally has one electron in the $3d$ shell, is
potent for strong correlation physics and is possibly another
realization of an effective one-band model beyond the cuprates. Here
monolayer VS$_2$ in both the trigonal prismatic and octahedral phases
is investigated using density functional theory plus Hubbard $U$
(DFT+$U$) calculations. Trigonal prismatic VS$_2$ has an isolated
low-energy band that emerges from a confluence of crystal field
splitting and direct V--V hopping. Within spin density functional
theory, ferromagnetism splits the isolated band of the trigonal
prismatic structure, leading to a low-band-gap $S=1/2$ ferromagnetic
Stoner insulator; the octahedral phase is higher in energy. Including
the on-site interaction $U$ increases the band gap, leads to Mott
insulating behavior, and for sufficiently high values stabilizes the
ferromagnetic octahedral phase. The validity of DFT and DFT+$U$ for
these two-dimensional materials with potential for strong electronic
correlations is discussed. A clear benchmark is given by examining the
experimentally observed charge density wave (CDW) in octahedral
VS$_2$, for which DFT grossly overestimates the bond length
differences compared to known experiments; the presence of CDWs is
also probed for the trigonal prismatic phase. Finally, we investigate
why only the octahedral phase has been observed in experiments and
discuss the possibility of realizing the trigonal prismatic phase. Our
work suggests trigonal prismatic VS$_2$ is a promising candidate for
strongly correlated electron physics that, if realized, could be
experimentally probed in an unprecedented fashion due to its monolayer
nature.

\end{abstract}

\date{\today}
\maketitle

\section{Introduction}

Transition metal dichalcogenides (TMDCs), composed of layers of
chalcogen--metal--chalcogen units (hereto called monolayers) that
stack and adhere via weak bonding, are a diverse class of materials
known to exhibit charge density waves, metal-insulator transitions,
superconductivity, and novel optoelectronic
properties.\cite{wilson_transition_1969} Recent breakthroughs in the
ability to isolate and manipulate few-layer and monolayer materials,
derived from TMDCs like MoS$_2$ and other layered crystals such as
graphite, have enabled new possibilities for device applications as
well as fundamental studies of low-dimensional
systems.\cite{novoselov_two-dimensional_2005}

Many TMDCs are nominally $d^0$ (e.g. TiS$_2$) or band insulators in
which an even number of $d$ electrons completely fills the valence
band (e.g. MoS$_2$). Such configurations preclude the possibility of
strong electronic correlations and/or magnetism in the ground state.
However, there are known examples from experiments of non-oxide
layered materials exhibiting magnetism and in some cases insulating
behavior. Spin-3/2 Cr$X$Te$_3$ is a ferromagnetic insulator with Curie
temperature of 33 K for $X$=Si and 61 K for $X$=Ge; monolayers in this
class of materials have been predicted to be stable with ferromagnetic
exchange as
well.\cite{carteaux_magnetic_1991,carteaux_crystallographic_1995,li_crxte3_2014,sivadas_magnetic_2015,zhuang_computational_2015}
The spin-1/2 insulator Cr$X_3$ is a ferromagnet below 37 K for $X$=Br
and 61 K for $X$=I; in CrCl$_3$ ferromagnetic layers stack in an
antiferromagnetic pattern with a N\'{e}el temperature of 17
K.\cite{de_haas_further_1940,tsubokawa_magnetic_1960,cable_neutron_1961,dillon-jr_magnetization_1965}
Ferromagnetic Fe$_3$GeTe$_2$, which is metallic, has a substantial
Curie temperature of 150
K.\cite{deiseroth_fe3gete2_2006,chen_magnetic_2013} In-plane
antiferromagnetism is also observed; MnPS$_3$ and MnPSe$_3$ are
spin-5/2 antiferromagnets with N\'{e}el temperatures of 78 and 74 K,
respectively.\cite{wildes_spin_1998,jeevanandam_magnetism_1999}
Additionally, there are numerous antiferromagnets in the family of Fe
pnictide superconductors.\cite{dai_antiferromagnetic_2015}


VS$_2$ is an interesting candidate among the many possible TMDCs. Here
nominal electron counting indicates that V donates two electrons to
each S, leaving it in a $d^1$ (i.e., spin-1/2)
configuration. Therefore, VS$_2$ might be potent for strong electronic
correlation physics, especially since its $3d$ electrons will be
significantly more localized than the $4d$ or $5d$ electrons of
NbS$_2$ or TaS$_2$, respectively. Similarly, the electronic states of
the sulfur anion should be more localized than those of selenium or
tellurium.


The structure of a monolayer TMDC consists of one metal layer
sandwiched between two chalcogen layers with each layer corresponding
to a triangular lattice. This gives rise to two basic types of
chalcogen-metal-chalcogen stacking: ABA stacking, in which the metal
layer hosts a mirror plane, or ABC stacking. The latter gives rise to
approximate octahedral coordination of the transition metal (TM) by
chalcogens, which results in the five-fold $d$ manifold splitting into
a 3-fold set ($T_{2g}$) and a 2-fold set ($E_g$) of orbitals. More
precisely, the octahedral environment experiences a trigonal
distortion due to the ability of the chalcogens to relax in the
out-of-plane direction. This results in a point group symmetry
lowering $O_h\rightarrow D_{3d}$ and a further splitting of the $d$
orbitals $T_{2g} \rightarrow A_{1g}+E_g'$. For convenience, we refer
to the distorted octahedral ($D_{3d}$) phase as the OCT phase in the
remainder of this paper.

\begin{figure}[t]
\begin{center}
\includegraphics[width=\linewidth]{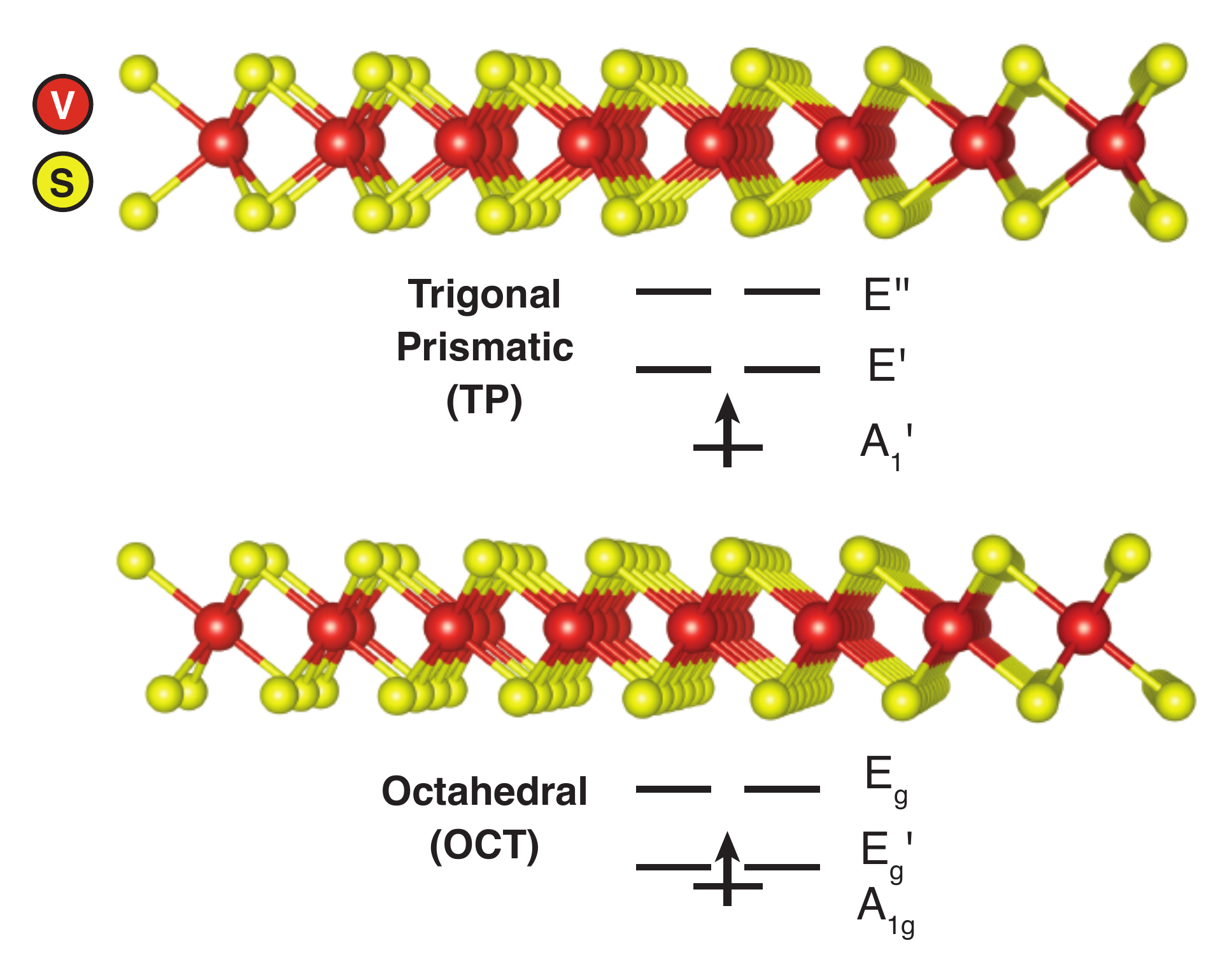}
\end{center}
\caption{Side view of crystal structures of trigonal prismatic and
  octahedral monolayer VS$_2$ and schematic V 3$d$ orbital fillings
  from crystal field theory. The red and yellow spheres represent
  ionic positions of V and S,
  respectively.\label{structures_level_diagrams}}
\end{figure}

Alternatively, ABA stacking results in a trigonal prismatic (TP)
coordination of the TM by the chalcogens. The TP coordination, which
is compared to that of the OCT structure in
Fig. \ref{structures_level_diagrams}, splits the $d$ manifold into a
one-fold $A_1'$ orbital and two different types of two-fold orbitals
($E'$ and $E''$). Both OCT and TP coordinations are possible for
VS$_2$, and the TP coordination is particularly intriguing since it
could potentially be a physical realization of a one-band model with
strong interactions; this rare feature is a hallmark of the copper
oxide (cuprate) high-temperature
superconductors.\cite{zhang_effective_1988}


Experimentally the TP phase has not been realized, but bulk VS$_2$ was
first synthesized in the OCT phase in the 1970s by deintercalating
LiVS$_2$.\cite{murphy_preparation_1977} It exhibits a charge density
wave (CDW) below $T=305$ K with a wavevector $q\approx2/3\ K$, where
$K$ is the corner of the Brillouin
zone.\cite{murphy_preparation_1977,tsuda_51v_1983,mulazzi_absence_2010}
In the CDW phase Mulazzi \textit{et al.} found metallic resistivity
and no lower Hubbard band in the photoemission spectrum, suggesting
rather weak electronic correlations.\cite{mulazzi_absence_2010} Only a
very small paramagnetic response was observed in the magnetic
susceptibility, which it was suggested might stem from V located in
between neighboring VS$_2$ monolayers. A more recent high-pressure
synthesis by Gauzzi \textit{et al.} found much more appreciable local
magnetic moments but no long-range CDW, and it was speculated that
``nm-size domains'' might be responsible.\cite{gauzzi_possible_2014}
Using phonon calculations, they also showed that the presence of a CDW
soft mode is very sensitive to the lattice parameters. Nanosheets,
though not a monolayer, of OCT VS$_2$ have been synthesized and
interpreted as showing
ferromagnetism.\cite{feng_metallic_2011,feng_giant_2012,gao_ferromagnetism_2013,zhong_ferromagnetism_2014}


Here we employ first-principles electronic structure calculations
based on DFT to explore the physics of VS$_2$. We focus on a single
layer of the material since the realization of a strongly correlated
monolayer material could enable one to probe Mott physics via gating
and strain in an unprecedented way. We find that DFT captures the
$q=2/3\ K$ CDW in OCT VS$_2$ and explains the lack of correlations
observed experimentally, though it substantially overestimates the
structural distortion. The addition of an appreciable on-site Hubbard
$U$ interaction to the V site leads to anti-aligned spins in OCT
VS$_2$ and yields V--V bond length distortions and metallic behavior
in reasonable agreement with known experiments. Unlike the OCT phase,
we find that TP VS$_2$ has an isolated low-energy $A_1'$ band at the
level of non-spin-polarized DFT due to the crystal field and direct
V--V hopping. The preferred magnetic order is ferromagnetic, as
opposed to the antiferromagnetic ordering found in the cuprates, and
this magnetism opens up a small band gap by splitting the $A_1'$
band. The on-site interaction leads to a low-band-gap $S=1/2$
ferromagnetic Mott insulator. For a narrow range of $U$ we find
evidence for a CDW in TP VS$_2$. Although DFT predicts ferromagnetic
TP VS$_2$ is the ground state, for moderate values of $U$ we find the
OCT structure becomes thermodynamically favored.

\section{Computational Details}

Density functional theory (DFT)
\cite{hohenberg_inhomogeneous_1964,kohn_self-consistent_1965}
calculations within the generalized gradient approximation of Perdew,
Burke, and Ernzerhof \cite{perdew_generalized_1996} are performed
using the Vienna \textit{ab initio} simulation package
(\textsc{vasp}).\cite{kresse_ab_1994,kresse_ab_1993,kresse_efficient_1996,kresse_efficiency_1996}
The Kohn-Sham equations are solved using a plane-wave basis set with a
kinetic energy cutoff of 500 eV and the projector augmented wave
method.\cite{blochl_projector_1994,kresse_ultrasoft_1999} The
out-of-plane lattice vector length is chosen to be 20\ \AA. To sample
reciprocal space we employ a $24\times24\times1$ $k$-point grid for
the primitive unit cell and $k$-point grids with approximately the
same $k$-point density for supercells. We utilize the tetrahedron
method with Bl\"{o}chl corrections\cite{blochl_improved_1994} for all
calculations except for structural relaxations and phonon calculations
in metals, for which we employ the first-order Methfessel-Paxton
method\cite{methfessel_high-precision_1989} with a 50 meV
smearing. The total energy, ionic forces, and stress tensor components
are converged to 10$^{-6}$ eV, 0.01 eV/\AA, and 10$^{-3}$ GPa,
respectively.

To compute maximally-localized Wannier functions (MLWF) we employ the
\textsc{wannier90} code.\cite{mostofi_wannier90:_2008} The
rotationally-invariant DFT+$U$ approach with fully localized limit
(FLL) double counting\cite{liechtenstein_density-functional_1995} is
used to explore the impact of an on-site Hubbard $U$ on the V 3$d$
electrons. Values of on-site Coulomb repulsion $U$ are computed from
first principles via the linear response approach of Cococcioni and de
Gironcoli.\cite{cococcioni_linear_2005} We do not employ an on-site
exchange interaction $J$ since this effect is present within spin
density functional theory.\cite{park_density_2015} We use the direct
(supercell) approach in \textsc{phonopy}\cite{phonopy} to compute
phonon dispersion relations. For these calculations we employ a
5$\times$5$\times$1 supercell for smaller $U$ and a larger
6$\times$6$\times$1 supercell for $U>$ 3 eV, which we find is needed
to capture the presence of soft mode instabilities. Phonons at select
$q$-points are obtained using the frozen phonon method to assess
supercell convergence of direct calculations. Images of crystal
structures are generated with \textsc{vesta}.\cite{momma2011vesta}

\section{Results and Discussion}

\subsection{CDW in OCT VS$_2$ within DFT}\label{oct_dft_cdw}
Given that a collection of experiments exist for the bulk OCT phase,
we begin by addressing the physics of the OCT monolayer. Since bulk
OCT VS$_2$ is known to undergo a CDW transition below $T=305$
K,\cite{murphy_preparation_1977,tsuda_51v_1983,mulazzi_absence_2010}
we explore the presence of such a CDW in the monolayer OCT
structure. We compute the phonon frequencies using the frozen phonon
method for $q=2/3\ K$, the experimental CDW wavevector from electron
microscopy,\cite{mulazzi_absence_2010} and verify the soft mode in the
non-spin-polarized (NSP) bulk OCT phase as found in a previous
study.\cite{gauzzi_possible_2014} We find the frequency is
$\omega=60i$ cm$^{-1}$. For the monolayer, at this wavevector we find
the same soft mode in the NSP state now with a slightly softer
frequency $\omega=80i$ cm$^{-1}$. Given the experimental CDW
wavevector is in-plane and the similarity of the soft mode for the
bulk and the monolayer, we expect the monolayer CDW to be
representative of that of the bulk. Additionally, at a slightly
different wavevector of $q=3/5\ K$ we find a soft mode of smaller
magnitude $\omega=48i$ cm$^{-1}$ in the monolayer.

\begin{figure}[tbp]
\includegraphics[width=\linewidth]{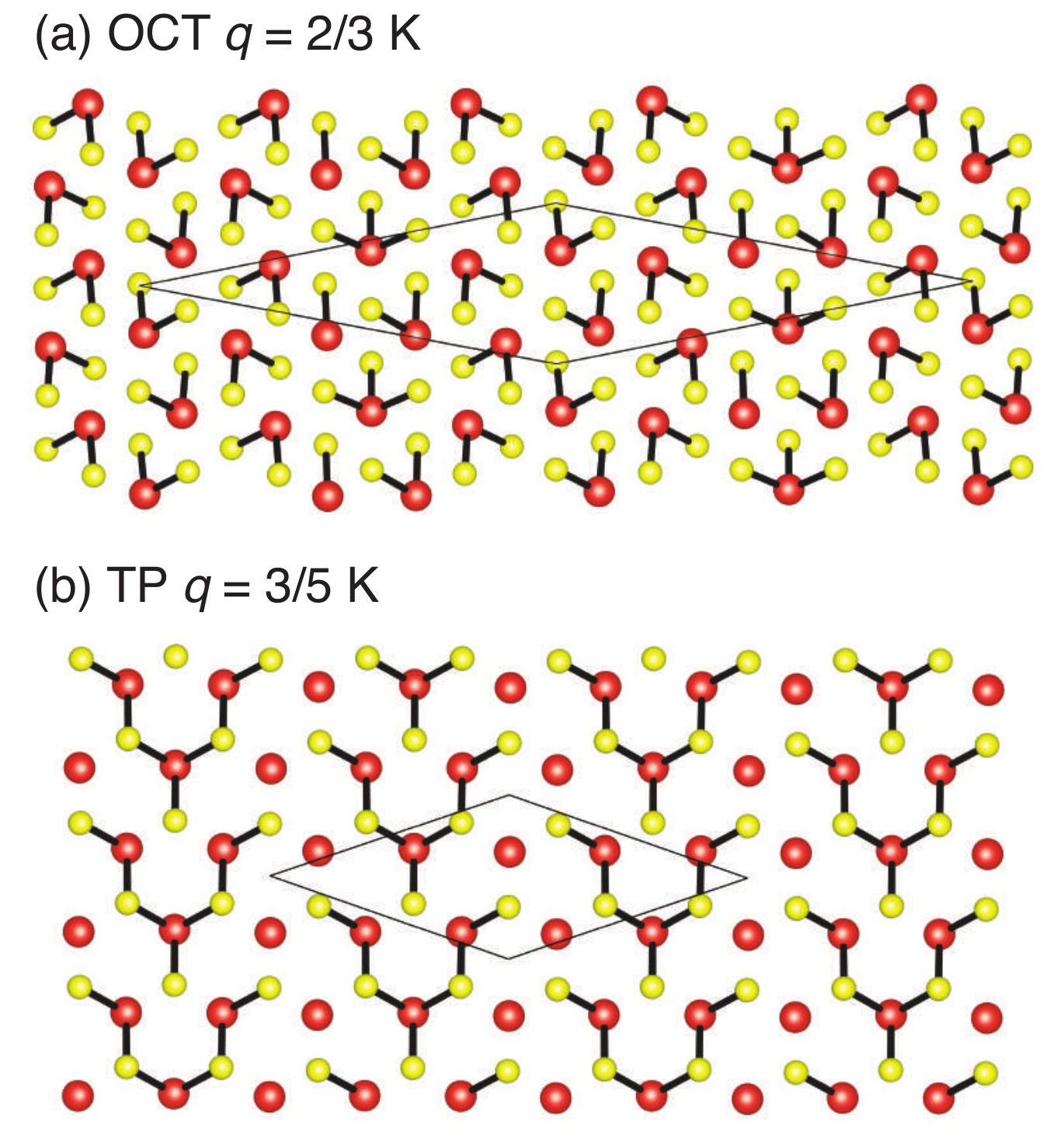}
\caption{Orthographic projection along the out-of-plane axis of the
  (a) FM $U=0$ $q=2/3\ K$ OCT and (b) FM $U=3.8$ eV $q=3/5\ K$ TP
  relaxed structures. Vanadium (sulfur) ions are indicated by red
  (yellow) spheres and the thick black lines show the shortest V--S
  bonds. The unit cell is indicated by thin black
  lines.\label{cdw_dists}}
\end{figure}

Without any CDW the lowest-energy state of monolayer OCT VS$_2$ is a
ferromagnetic (FM) metal with a V magnetic moment of 0.5 $\mu_B$,
which is 13 meV lower in energy than the NSP state. The relaxed NSP
$q=2/3\ K$ OCT CDW state is 12 meV lower in energy than the pristine
(without-CDW) FM state. Although we find no soft mode for the pristine
OCT FM structure, performing a further structural relaxation of the
NSP $q=2/3\ K$ OCT CDW structure with FM initialization leads to an
additional small (\textless1 meV) energy lowering (see
Fig. \ref{dft_energy_levels}). In this structure, depicted in
Fig. \ref{cdw_dists}(a), distinct V sites have one, two, or three
nearest-neighbor S atoms instead of the six of the pristine OCT
structure. The CDW has substantially suppressed the V magnetic moments
to 0.0--0.2 $\mu_B$, which is consistent with the weak correlations
observed by Mulazzi \textit{et al.} However, the V--S and V--V bond
lengths exhibit massive variations of 2.2--2.6 and 3.0--3.7 \AA\ ,
respectively. The range of V--V bond lengths deduced by Sun \textit{et
  al.} via x-ray absorption fine spectroscopy (XAFS) is only 0.19
\AA.\cite{sun_in-situ_2015} Therefore, DFT is severely overestimating
the structural deformation in the CDW state and beyond-DFT approaches
will be necessary to describe the OCT CDW phase; we address this point
in detail using DFT+$U$ in Sec. \ref{impact_U}. Also, additional
experimental studies would be helpful to understand the lack of
long-range CDW found using high-pressure synthesis.


\subsection{Non-spin-polarized DFT electronic structure}

\begin{figure}[b]
\begin{center}
\includegraphics[width=\linewidth]{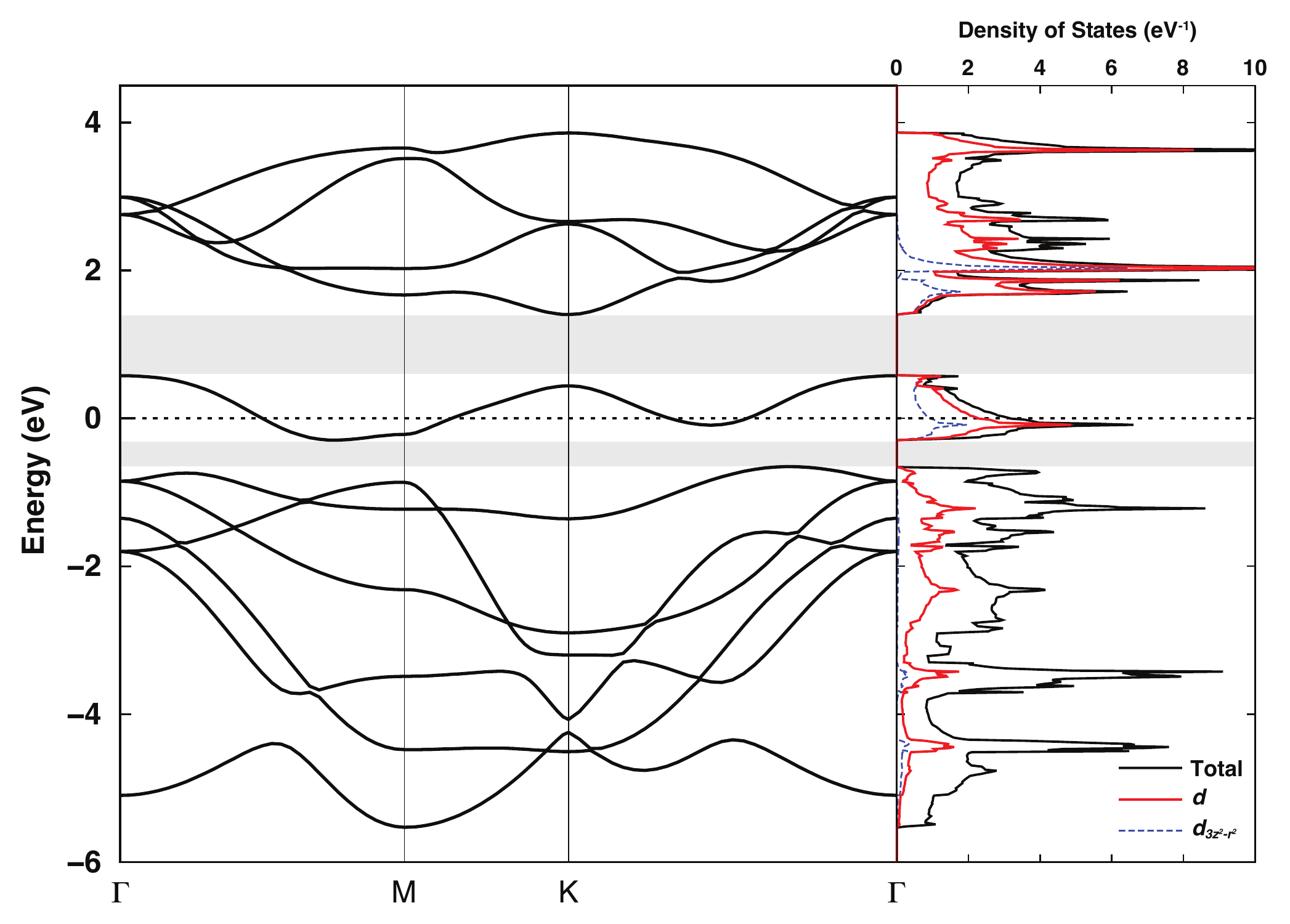}
\end{center}
\caption{NSP electronic band structure and total (solid black line),
  $d$ (solid red line), and $d_{3z^2-r^2}$ (dashed blue line) density
  of states for TP VS$_2$ within DFT. The black dotted line indicates
  the Fermi energy and the shaded areas illustrate the gaps around the
  isolated low-energy band. The $k$-point labels $\Gamma$, $M$, and
  $K$ correspond to the center, edge midpoint, and corner of the
  Brillouin zone.\label{tp_dft_bands_dos}}
\end{figure}

The NSP band structure and density of states for TP VS$_2$ are shown
in Fig. \ref{tp_dft_bands_dos}. We do find an isolated low-energy band
like in the crystal field picture shown in the top panel of
Fig. \ref{structures_level_diagrams}, but there is a major difference
with the simple schematic. The projected density of states shows this
isolated band is mainly of $d$ character, while the unoccupied
manifold above it has slightly less predominant $d$ character (i.e.,
stronger hybridization with S $p$); the manifold below is
predominantly S $p$ with some hybridization with V $d$. However,
projecting the V $d$ density of states onto just the $A_1'$ orbital
($d_{3z^2-r^2}$) reveals the main discrepancy with the simple
schematic: the isolated band is only roughly half $A_1'$ character and
the remaining half is $E'$ character. This puzzle was first noted by
Kertesz and Hoffman in the context of TMDCs several decades
ago.\cite{kertesz_octahedral_1984}


\begin{figure*}[t]
\includegraphics[width=\textwidth]{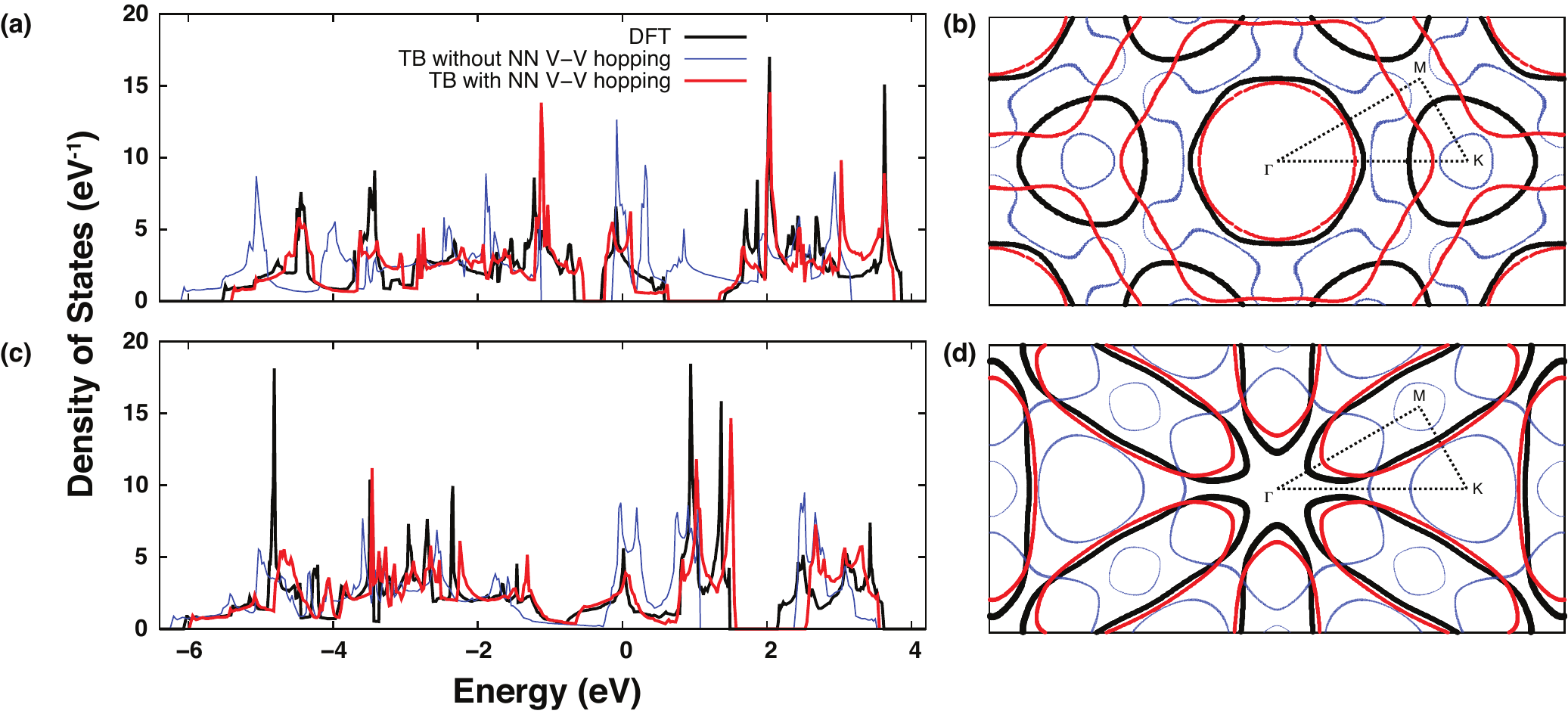}
\caption{(a) Density of states and (b) Fermi surface for NSP TP
  VS$_2$. The thick black lines correspond to DFT, while the thick red
  (thin blue) lines indicate tight binding results with (without) NN
  V--V hopping matrix elements. The dotted lines show the irreducible
  Brillouin zone. Corresponding plots for OCT VS$_2$ are shown in
  panels (c) and (d).\label{tb}}
\end{figure*}

In order to resolve this anomaly and to gain further insight into the
electronic structure of the TP phase, we compute MLWF for the full
$p$-$d$ manifold of TP VS$_2$, which results in atom-centered V
$d$-like and S $p$-like orbitals. The Hamiltonian is represented in
the MLWF basis, and we explore the impact of removing various matrix
elements in the Hamiltonian corresponding to V--S and V--V hoppings;
S--S hoppings are always retained. A similar analysis is performed for
the OCT phase for comparison.

Panels (a) and (c) of Fig. \ref{tb} show the density of states from
the MLWF Hamiltonian for NSP TP and OCT VS$_2$ (black curves),
respectively, which are identical to those of DFT by construction. The
OCT structure, unlike the TP structure, does not have an isolated
low-energy band since the crystal field splitting of the $T_{2g}$ into
$A_{1g}$ and $E_g'$ is relatively weak as is also typical for oxides
in this structure. Now we examine the tight binding (TB) approximation
in which we remove all V--S and V--V matrix elements beyond nearest
neighbor (NN) (thick red lines). In both phases, we qualitatively
reproduce all of the gaps and other prominent features of the
spectra. For both structures, we find V--V hopping beyond NN is
negligible, and therefore all of the quantitative deviation between
the black and the red curves is due to V--S hopping beyond NN.

If we only include NN V--S hoppings and no NN V--V hoppings (thin blue
lines) one still captures the qualitative features of the spectra for
the OCT structure, though there are now large quantitative
differences. However, for TP phase there is a qualitative change:
there is no longer a gap between the isolated $d$ band and the
higher-energy $d$ bands. Therefore, the V--V hopping plays a strong
contribution in splitting off the isolated band. Furthermore, it
addresses the observation presented by Kertesz and Hoffman. The fact
that the NN V--V hoppings have a strong interorbital component
explains why $A_1'$ only contributes half of character of the isolated
band. Interestingly, we also find that the rapid decay of these direct
TM--TM hoppings with strain explains the semiconductor-to-semimetal
transition in the isostructural $d^2$ material MoS$_2$ under
strain.\cite{scalise_strain-induced_2012}



Panels (b) and (d) of Fig. \ref{tb} illustrate the Fermi surfaces of
the TP phase and OCT phase, respectively. In DFT, the Fermi surface of
the TP structure has hole pockets centered at $\Gamma$ and $K$, while
that of the OCT structure has a single cigar-shaped electron pocket
centered at $M$. For the OCT structure the TB approximation is
sufficient to properly capture the Fermi surface topology, but for the
TP structure this is not the case and longer-range V--S hopping is
needed.

At this level of theory we predict an isolated low-energy band in the
TP phase, but as discussed in the next section there is a
ferromagnetic instability once spin polarization is included even at
the DFT level. This strongly suggests electronic correlations will be
important in the TP phase of this material, which therefore is our
focus for the remainder of this paper.

\subsection{DFT energy level diagram}

\begin{figure}[tbp]
\includegraphics[width=\linewidth]{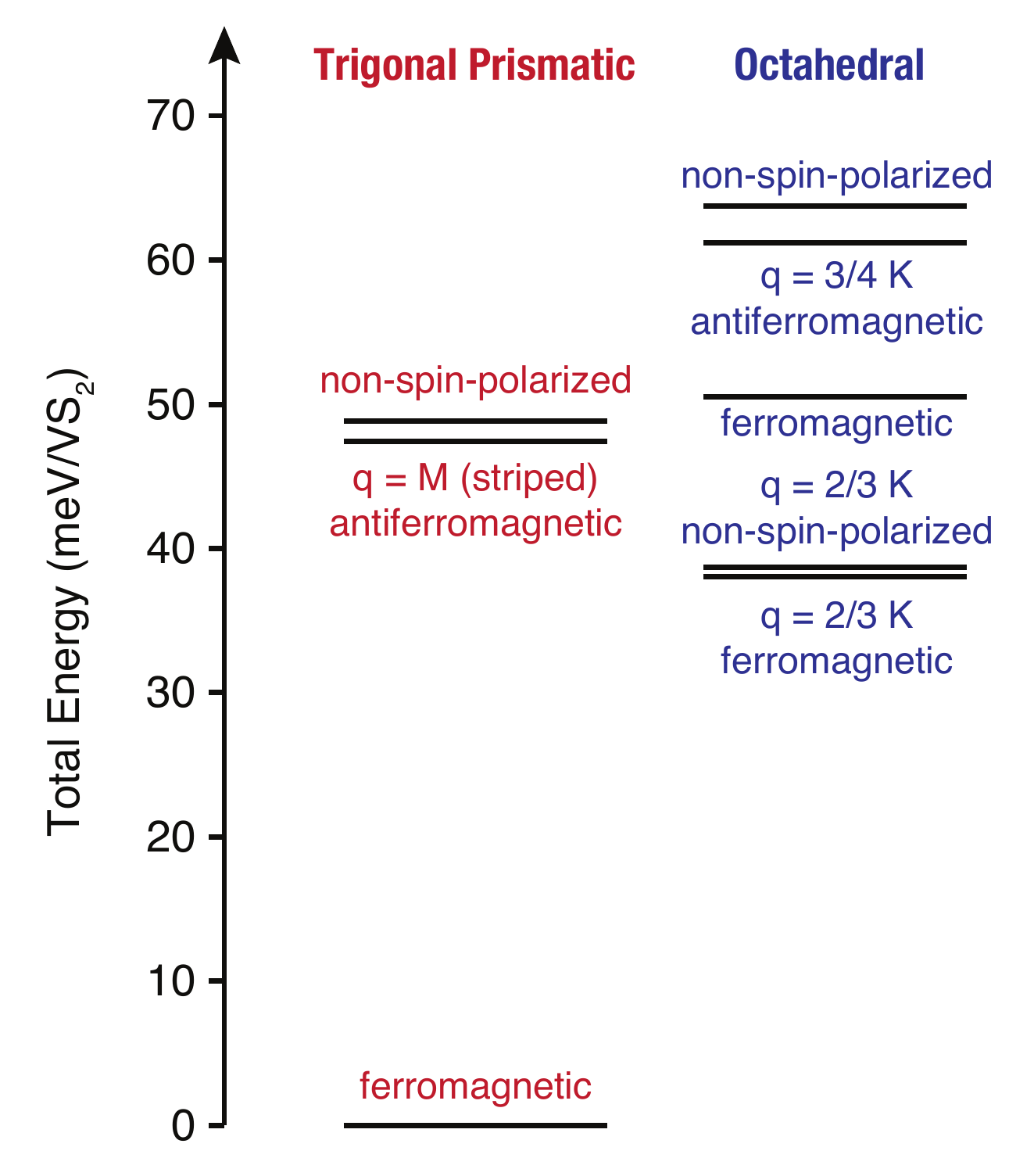}
\caption{Energy level diagram for TP (left, in red) and OCT (right, in
  blue) VS$_2$ within DFT. The energy of the FM TP state is used as a
  reference energy.\label{dft_energy_levels}}
\end{figure}

The total energy of different structures and magnetic configurations
of monolayer VS$_2$ within DFT is shown in
Fig. \ref{dft_energy_levels}. For the NSP states, the TP structure is
lower in energy than the OCT structure by 15 meV. For both structures,
the formation of a FM state results in a significant energy lowering
compared to the NSP state. The magnitude of the energy decrease is 13
meV for OCT and 49 meV for TP. In the FM state, V in the TP structure
is fully spin polarized with a magnetic moment of 1.0 $\mu_B$ whereas
for the OCT structure the moment is only 0.5 $\mu_B$, indicating that
the TP phase exhibits stronger signatures of electronic
correlations. For the OCT phase one must also consider the CDW phase,
which lowers the OCT energy by 12 meV compared to the FM state and
greatly weakens the magnetism giving moments of only 0.0--0.2
$\mu_B$. Ultimately, the TP FM state is the ground state since it is
still far lower in energy (38 meV) than the OCT FM CDW phase. The only
remaining task is to provide evidence that there are no other magnetic
or phonon instabilities.

To confirm the exchange is FM in VS$_2$, we also investigate $q=M$ and
$q=3/4\ K$ antiferromagnetic (AFM) configurations. For the TP phase,
only the striped ($q=M$) AFM configuration is found to converge. This
metastable state is metallic with small V magnetic moments of $\pm$0.2
$\mu_B$ and is only 1.4 meV lower in energy than the NSP
state. Therefore, TP VS$_2$ strongly prefers ferromagnetism and we
interpret it as a ``Stoner insulator'' rather than a Mott insulator at
the level of spin-dependent DFT, given that a gap does not persist for
an arbitrary magnetic ordering. For the OCT structure a metastable
$q=3/4\ K$ AFM configuration is found only 2.4 meV lower in energy
than the NSP state, and it similarly is metallic with small V moments
of $\pm$0.4 $\mu_B$. The FM nature of the exchange in this system is
not unexpected since the V--S--V angle is 84--85 degrees, close to the
90-degree ferromagnetism given by the Goodenough-Kanamori
rules.\cite{goodenough_theory_1955,goodenough_interpretation_1958,kanamori_superexchange_1959}


\begin{figure}[tbp]
\includegraphics[width=\linewidth]{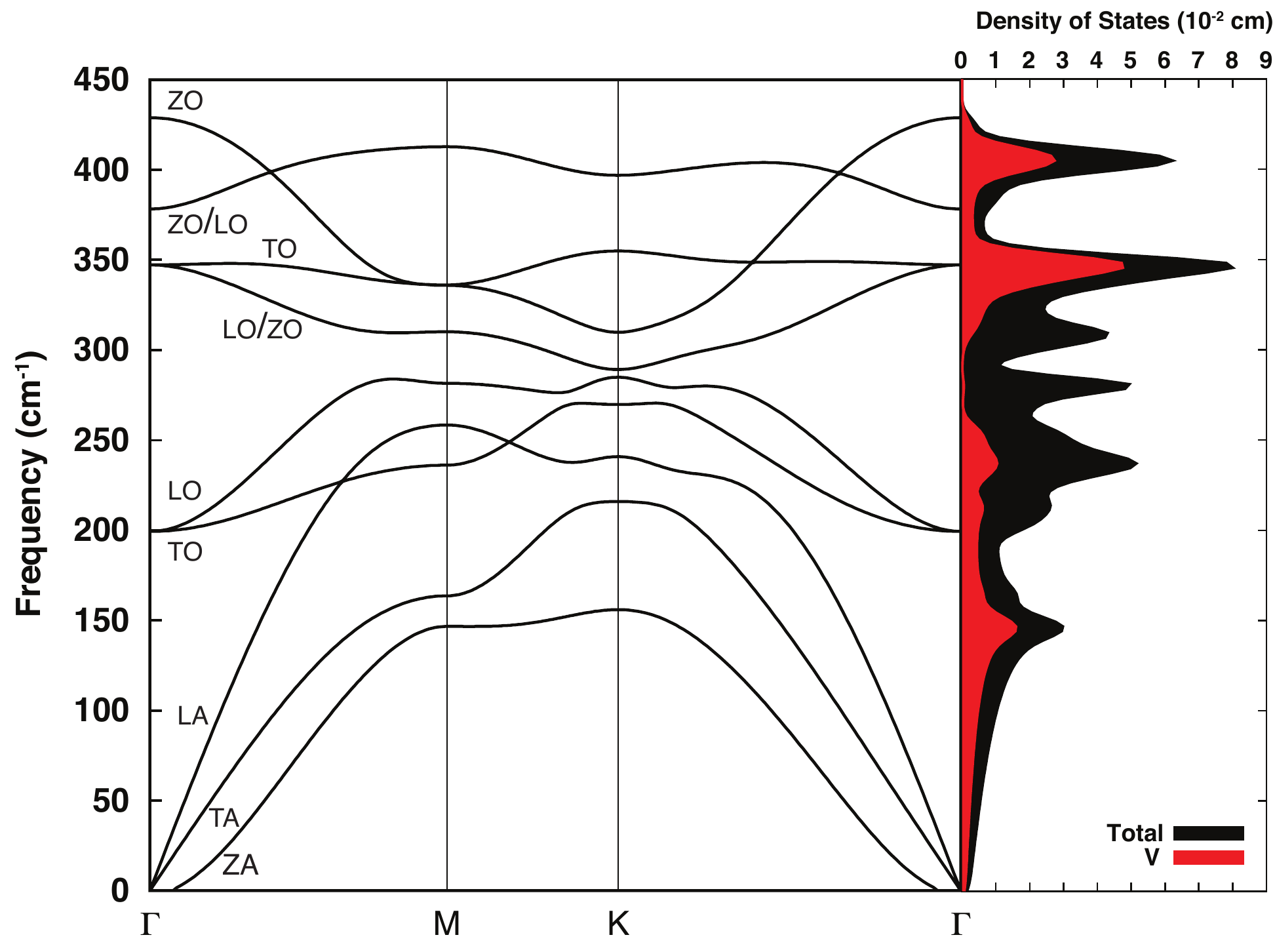}
\caption{Phonon dispersion relation and total (black) and V-projected
  (red) phonon density of states for FM TP VS$_2$ within DFT. The band
  labels identify the mode character near the $\Gamma$
  point. \textsc{z}, \textsc{t}, \textsc{l}, \textsc{a} and \textsc{o}
  refer to out-of-plane, transverse, longitudinal, acoustic, and
  optical branches, respectively.\label{dft_phonons}}
\end{figure}

We compute the phonon dispersion and density of states of FM TP
VS$_2$, shown in Fig. \ref{dft_phonons}, to assess the dynamic
stability of this phase. The out-of-plane acoustic (\textsc{za})
branch has the $\omega \sim q^2$ form near $\Gamma$ characteristic of
two-dimensional materials. There is no frequency gap between the
acoustic and optical branches. The out-of-plane optical (\textsc{zo})
branches are the highest-frequency phonons. Since there are no modes
with imaginary frequency, this phase is stable at the level of DFT.

The above analysis of the magnetism and the phonons allows us to
conclude that the FM TP phase is the ground state within DFT. One
would not interpret this as a Mott insulator within DFT given that the
band gap does not persist for all spin configurations.

\subsection{Impact of on-site Hubbard $U$}
\label{impact_U}

We use the linear response approach\cite{cococcioni_linear_2005} to
estimate the correlation strength $U$ for V in VS$_2$. Computing
screened interactions for use in beyond-DFT methods is still an active
area of research, but the linear response approach is useful to set a
baseline for the expected value of $U$. For FM states, we obtain
$U=3.84$ eV for the TP phase and $U=3.99$ eV for OCT phase. For the TP
phase, we also compute the $U$ for the NSP state and obtain 4.14
eV. These values are generally smaller than oxides of
vanadium\cite{xu_accurate_2015} and larger than sulfides of titanium
and
tantalum.\cite{sanchez_electronic_2008,darancet_three-dimensional_2014}
Ultimately, one still needs to carefully investigate the effect of $U$
on the physical observables given the methodological uncertainties.

\begin{table*}[h]
\begin{tabular}{ccccc}
\hline\hline
& $U$ (eV) & $\Delta$E (meV) & V--S bond length range (\AA) & V magnetic moment range ($\mu_B$)\\
\hline
NSP $q=3/5\ K$ & 0 & -17 & 2.22--2.52 & ---\\
& 1 & -20    & 2.23--2.51 & ---\\
& 2 & -33    & 2.24--2.51 & ---\\
& 3 & -60   & 2.25--2.51 & ---\\\hline
NSP $q=2/3\ K$ & 0 & -25    & 2.18--2.57 & ---\\
& 1 & -27    & 2.20--2.56 & ---\\
& 2 & -34    & 2.21--2.55 & ---\\
& 3 & -60   & 2.25--2.52 & ---\\\hline
FM $q=3/5\ K$ & 0 & -7 & 2.21--2.53 & 0.03--0.38\\
& 1 & -2    & 2.30--2.42 & 1.17--1.19\\
& 2 & -14   & 2.26--2.51 & 1.21--1.39\\
& 3 & -1   & 2.37--2.42 & 1.30--1.40\\\hline
FM $q=2/3\ K$ & 0 & -12    & 2.18--2.57 & -0.02--0.18\\
& 1 & -12    & 2.26--2.47 & 1.14--1.20\\
& 2 & -10    & 2.27--2.49 & 1.27--1.32\\
& 3 & -1   & 2.39--2.40 & 1.28--1.33\\\hline\hline
\end{tabular}
\begin{centering}
\caption{Total energy change per formula unit with respect to the
  pristine structure of the same magnetic state, V--S bond length
  range, and V magnetic moment range for the NSP and FM states of OCT
  VS$_2$ with $q=3/5\ K$ and $q=2/3\ K$ relaxed structures.
\label{oct_soft_mode_table}}
\end{centering}
\end{table*}

Another useful benchmark that could provide a bound for $U$ is the CDW
in the OCT phase. We performed structural relaxations to check if the
CDW is still captured for finite $U$. The total energy lowering
$\Delta$E, V--S bond length range, and V magnetic moment range for the
relaxed structures are given in Table \ref{oct_soft_mode_table} for
NSP and FM OCT VS$_2$ for $q=3/5\ K$ and $q=2/3\ K$. For the NSP
states the energy lowering from the CDW increases substantially with
$U$ and is 60 meV for $U=3$ eV. For the FM states, the CDW persists
for moderate values of $U$ but it is substantially dampened once $U$
is 3 eV with a total energy lowering of only 1 meV. However, at $U=3$
we find evidence for a new $q=2/3\ K$ CDW ground state with AFM-like
correlations. This system is a ferrimagnetic metal with 2 V moments of
1.3 $\mu_B$, 3 V moments of 1.4 $\mu_B$, and 4 V moments of -1.2
$\mu_B$. We refer to it as an AFM state for simplicity since the total
magnetization is only 0.21 $\mu_B$ per formula unit.

Further evidence for this tendency for AFM correlations in OCT VS$_2$
for larger $U$ comes from calculations of the $q=M$ and $q=3/4\ K$ AFM
states. For $U=3$ eV the $q=M$ and $q=3/4\ K$ AFM states are also
lower in energy than the pristine FM state by 29 and 19 meV,
respectively. The $q=2/3\ K$ AFM CDW state is even lower in energy, 39
meV lower than the pristine FM state, and therefore is the ground
state. For $U=4$ eV this trend persists as $q=M$ and $q=3/4\ K$ phases
with anti-aligned magnetic moments are lower in energy than the
pristine FM phase by 35 and 29 meV, respectively. It should be
emphasized that these anti-aligned magnetic states are strongly
coupled to the structural distortions; performing an unrelaxed $U=3$
eV calculation based on the FM $U=0$ or $U=3$ eV relaxed structure of
the primitive unit cell (i.e., without any CDW) demonstrates that the
FM spin ordering persists as the ground state.

%
%

\begin{figure}[tbp]
\includegraphics[width=\linewidth]{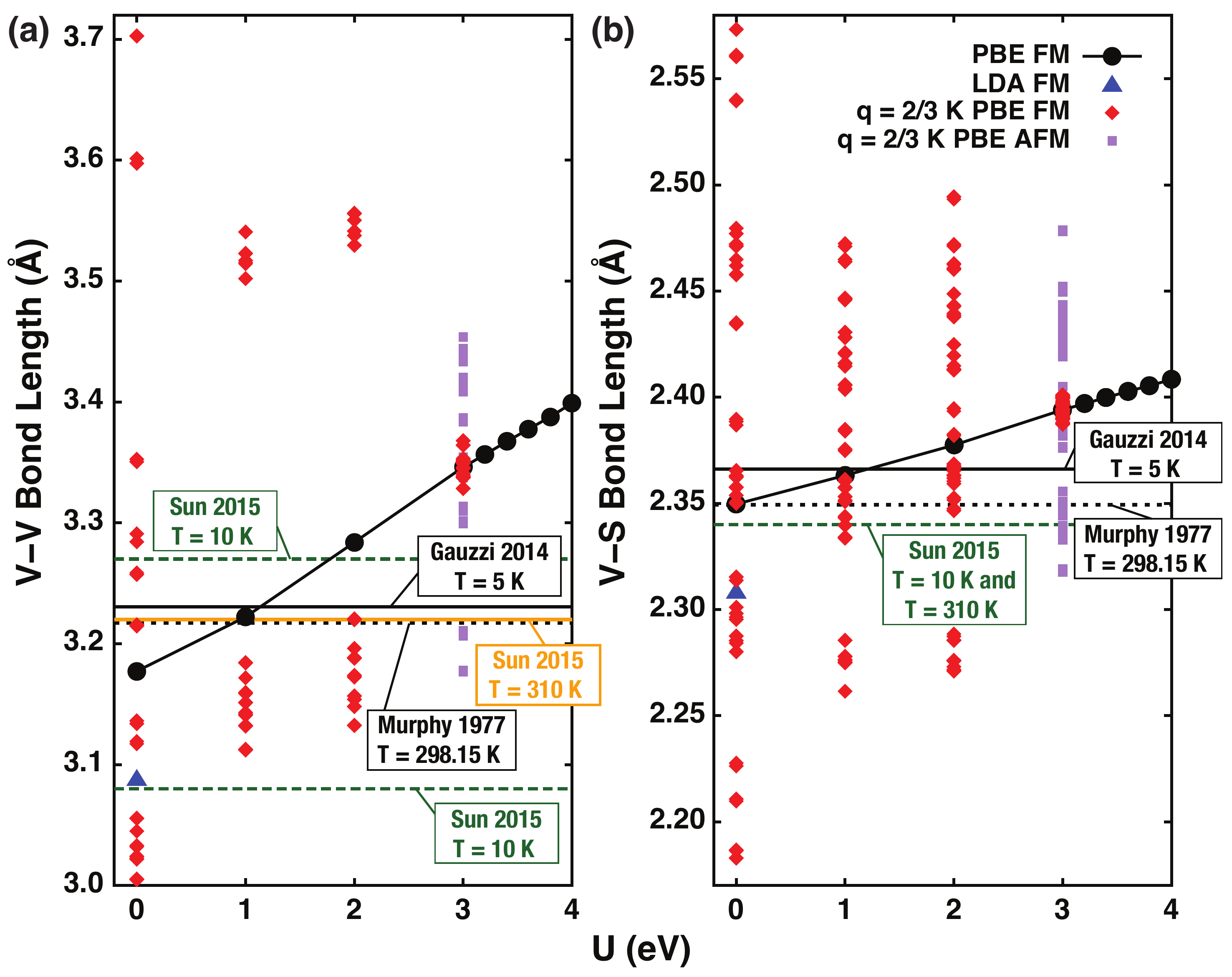}
\caption{ (a) V--V and (b) V--S bond lengths for OCT VS$_2$ in the
  pristine FM phase, $q=2/3\ K$ FM CDW phase, and $q=2/3\ K$ AFM CDW
  phase as a function of $U$. For comparison, the $U=0$ value for the
  pristine FM phase is also shown within the local density
  approximation (LDA).\label{oct_bond_lengths}}
\end{figure}

To assess which regime of $U$ best agrees with experiments on the CDW
phase, we compare the V--V and V--S bond lengths of our calculated
structures to those of known experiments in
Fig. \ref{oct_bond_lengths}. For the V--V bond length the
high-temperature value of Sun \textit{et al.} agrees well with that of
Murphy \textit{et al.}, which may be reasonable since the temperature
is approaching the CDW transition at 305 K.  Gauzzi \textit{et al.},
who do not find a long-range CDW, observe a slightly larger V--V bond
length at low temperature.  The work of Sun \textit{et al.} is the
only one that presents bond lengths at low temperature well within the
CDW phase; they report a V--V bond length difference of 0.19 \AA.


Applying DFT+$U$ while not allowing spontaneously broken translational
symmetry, the V--V and V--S bond lengths of the pristine FM state
increase roughly linearly with $U$. For this state, within DFT ($U=0$)
PBE predicts larger bond lengths than the local density approximation
(LDA) as is typical. As discussed in Sec. \ref{oct_dft_cdw}, for $U=0$
the range of V--V bond lengths of the $q=2/3\ K$ FM CDW phase (0.70
\AA) is over 3.5 times the low-temperature XAFS measurement from Sun
\textit{et al.} For $U=1$ and 2 eV the range we compute is smaller but
still over twice the experimental value, while the range collapses to
only 0.04 \AA for $U=3$ eV. Alternatively, reasonable agreement with
experiment occurs for the $U=3$ eV $q=2/3\ K$ AFM CDW phase. This
phase still contains an appreciable CDW distortion, unlike the
corresponding FM phase, and the range of V--V bond lengths of 0.28
\AA\ is comparable to that in experiment. Furthermore, the metallic
nature of this phase (unlike the gapped FM CDW phase) is qualitatively
consistent with the experimental
resistivity.\cite{murphy_preparation_1977,mulazzi_absence_2010,sun_in-situ_2015}
Therefore, an appreciable $U$ value of around 3 eV may be most
reasonable for OCT VS$_2$, and we find evidence for AFM correlations
in this regime. The V--S bond lengths show a similar trend: the $U=3$
eV $q=2/3\ K$ FM CDW phase exhibits a massive range of values for
$U=0$ that is dampened for $U=1$ and 2 eV and nearly disappears for
$U=3$ eV. We note that Sun \textit{et al.} reports only a single
temperature-independent V--S bond length, however. A detailed
structural refinement from experiment would be instrumental for a more
stringent evaluation of available first-principles methodologies.


DFT+$U$ corresponds to a Hartree-Fock (mean-field) solution to the
quantum impurity problem of dynamical mean-field
theory.\cite{georges_dynamical_1996,kotliar_electronic_2006} Given the
manner in which Hartree-Fock tends to overemphasize the effects of
interactions, it would not be surprising to require a smaller value of
$U$ relative to that of linear response to provide a proper
description. Especially given that there are currently no experiments
for the TP phase, the above analysis indicates the need to explore a
range of $U$ values in what follows.

\begin{figure}[tbp]
\includegraphics[width=\linewidth]{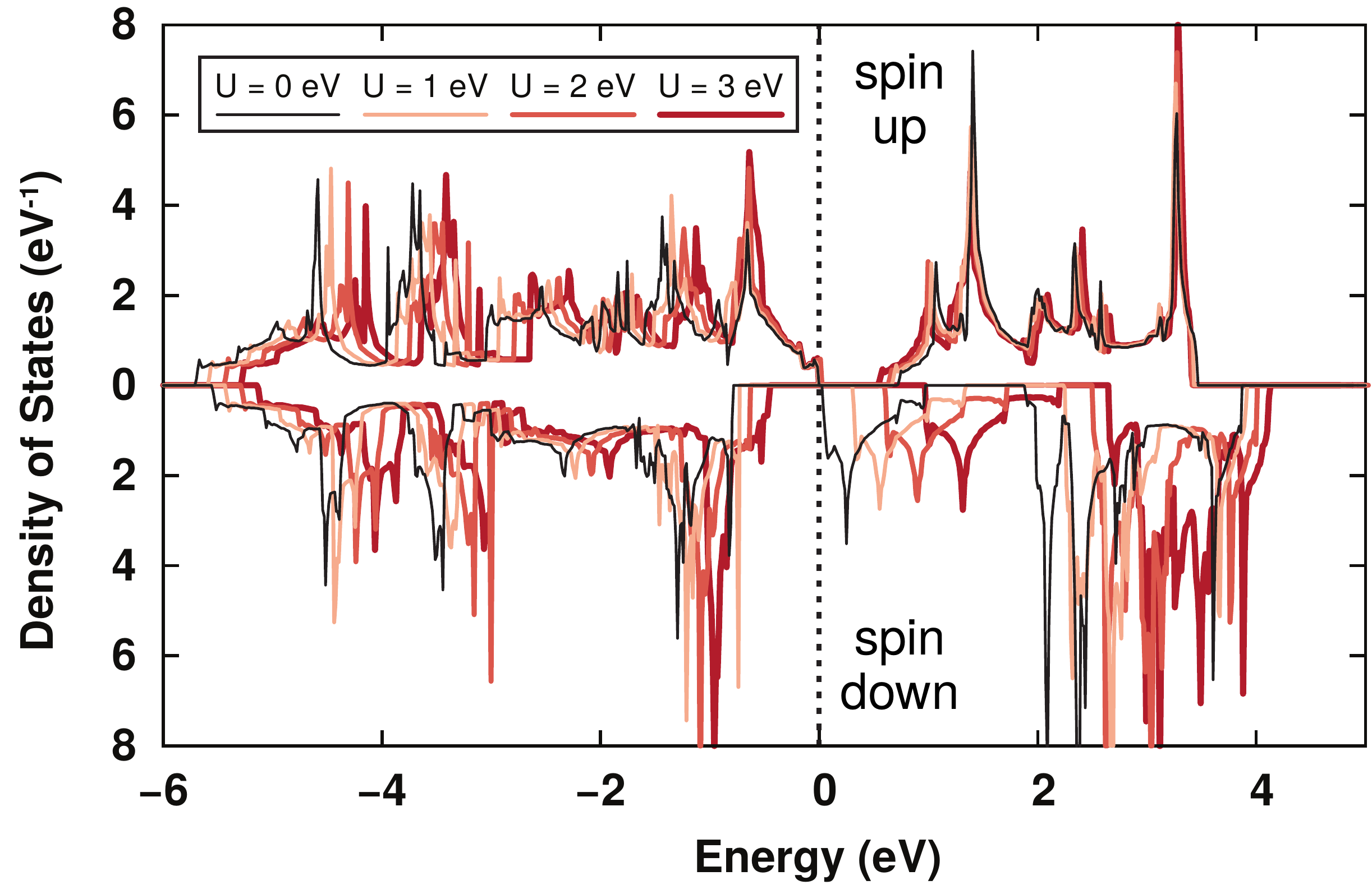}
\caption{ Electronic density of states for FM TP VS$_2$ for different
  values of $U$. The dotted black line indicates the valence band
  maximum.\label{dftu_dos}}
\end{figure}

We explore the effect of $U$ on the electronic spectrum of FM TP
VS$_2$ using DFT+$U$. As shown in Fig. \ref{dftu_dos}, for $U=0$
already there is a small band gap of 30 meV generated by the exchange
splitting of the $A_1'$ state. With increasing $U$ the spin-down
$A_1'$ state is shifted up in energy, which increases the band gap up
to 0.6 eV; the band gap saturates once the spin-up $E'$ levels become
the lowest unoccupied states. This value is somewhat smaller than the
1.1 eV band gap obtained via hybrid functional calculations, which is
presumably due to the nonlocality of the potential in the hybrid
functional.\cite{huang_prediction_2015} For small $U$, the $U$-induced
energy shift of correlated orbital $|d_\alpha\rangle $ with occupancy
$n_\alpha$ takes the form $U(1/2-n_\alpha)$ within DFT+$U$, so one
expects an occupied state ($n_\alpha=1$) to shift down in energy by
$U/2$ and an unoccupied state ($n_\alpha=0$) to shift up in energy by
$U/2$. In this case, however, the spin-up $d$ levels are significantly
hybridized such that their occupancies are very close to 1/2 (i.e.,
0.45--0.48) within DFT. This necessitates that the spin-up $d$
manifold is essentially fixed in energy for small $U$. The trend
happens to persist over the full range of $U$ shown, which is
responsible for the band gap saturation observed here as well as in a
previous study.\cite{huang_prediction_2015}

For $U$ of 2 and 4 eV the metastable striped $q=M$ AFM configuration
is 115 and 66 meV higher in energy than the FM state with a band gap
of 0.1 and 0.7 eV and V magnetic moments of $\pm$0.6 and $\pm$1.3
$\mu_B$, respectively. The insulating behavior for this higher-energy
magnetic configuration indicates that the system has been driven into
a regime of Mott physics, as crudely interpreted from DFT+$U$; this is
in contrast to the DFT description in terms of a Stoner instability.

\begin{figure}[tbp]
\includegraphics[width=\linewidth]{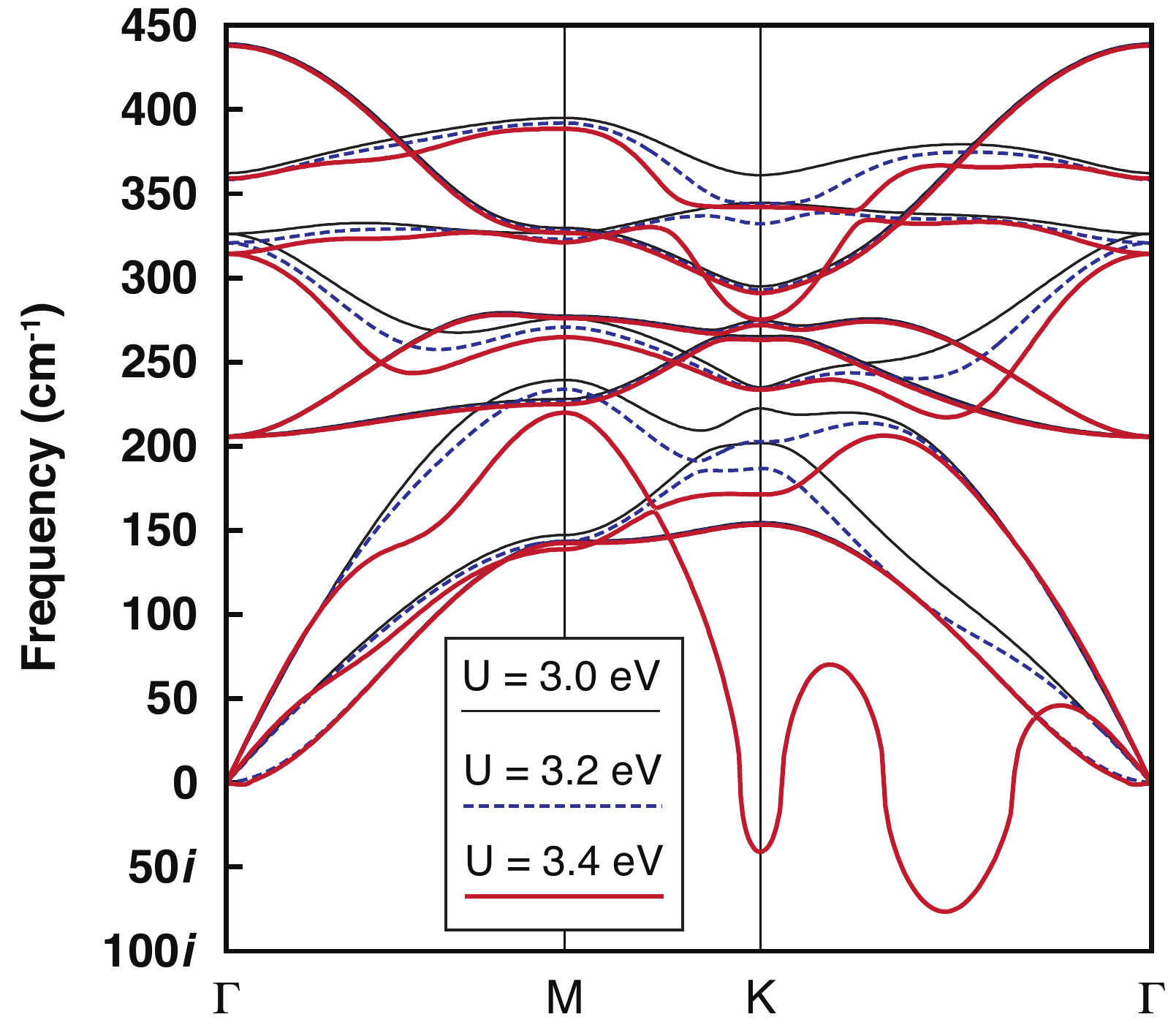}
\caption{Phonon dispersion relation for FM TP VS$_2$ for $U=3.0$ eV
  (thin solid black lines), $U=3.2$ eV (dashed thin blue lines), and
  $U=3.4$ eV (thick solid red line).\label{dftu_phonons}}
\end{figure}

We also examine the impact of $U$ on the phonon dispersion relation of
the FM TP state to assess the dynamical stability of VS$_2$. Figure
\ref{dftu_phonons} illustrates the main result. For $U=3.0$ eV the
phonons are all still stable, as in the DFT case. For $U=3.2$ eV one
can observe the formation of a small dip in the \textsc{ta} branch
between $\Gamma$ and $K$. Once $U$ is equal to 3.4 eV, a soft mode is
formed. There is an additional soft mode at $q=K$ whose eigenvalue is
smaller in magnitude.

\begin{table*}[h]
\begin{tabular}{ccccc}
\hline\hline
& $U$ (eV) & $\Delta$E (meV) & V--S bond length range (\AA) & V magnetic moment range ($\mu_B$)\\
\hline
& 3.4 & -0.1 & 2.38--2.40 & 1.38--1.39\\
& 3.6 & -3    & 2.38--2.40 & 1.41--1.41\\
$q=K$ & 3.8 & -10   & 2.38--2.41 & 1.44--1.44\\
& 4.0 & +9 & 2.38--2.43 & 1.32--1.50\\
& 4.2 & +16 & 2.38--2.44 & 1.33--1.52\\\hline
& 3.4 & -1    & 2.37--2.42 & 1.17--1.56\\
& 3.6 & -7    & 2.37--2.44 & 1.16--1.71\\
$q=3/5\ K$ & 3.8 & -19   & 2.36--2.45 & 1.18--1.82\\
& 4.0 & -34   & 2.36--2.46 & 1.19--1.90\\
& 4.2 & -45 & 2.36--2.47 & 1.20--1.97\\\hline\hline
\end{tabular}
\begin{centering}
\caption{Total energy change per formula unit, V--S bond length range,
  and V magnetic moment range for FM TP $q=K$ and $q=3/5\ K$ relaxed
  structures for several $U$ values.
\label{tp_soft_mode_table}}
\end{centering}
\end{table*}


To corroborate and refine our finding of $U$-induced soft modes in the
TP phase, we performed frozen phonon calculations at several
$q$-points. The frozen phonon method removes the possibility of image
interactions, which can cause errors in the supercell approach. For
$U=3.4$ eV we find a 130$i$ cm$^{-1}$ soft mode at the $K$ point, a
100$i$ cm$^{-1}$ soft mode at $q=1/2\ K$, and a 188$i$ cm$^{-1}$ soft
mode at $q=3/5\ K$; this reveals that the supercell approach is
qualitatively correct but with substantial quantitative errors.

We performed structural relaxations for the two wavevectors with the
softest phonon modes, $q=K$ and $q=3/5\ K$, using supercells
commensurate with those wavevectors. The total energy lowering
$\Delta$E, V--S bond length range, and V magnetic moment range for the
relaxed structures are given in Table \ref{tp_soft_mode_table}. For
$U=3.2$ eV no structural distortion is found for either
wavevector. With larger $U$ values, the relaxed structures exhibit
lower total energy and modulation of V--S bond lengths and V magnetic
moments. For $q=3/5\ K$ the magnitude of $\Delta$E increases
monotonically from 1 meV to 45 meV as $U$ increases, corresponding to
an enhanced CDW. The V--S bond lengths vary by as much as 0.09
\AA\ and the V magnetic moments differ by as much as 0.8 $\mu_B$ at a
given $U$. For 3.4 eV $\le U\le$ 3.8 eV the $q=K$ soft mode also shows
an appreciable but smaller energy lowering ($|\Delta E|\le 10$ meV)
with significantly smaller magnitudes of the differences in V--S bond
length (0.03 \AA) and V magnetic moment (0.01 $\mu_B$); for $U>3.8$ eV
this CDW state becomes higher in energy than the undistorted FM
state. For $U=5$ eV we do not find a stable (or even metastable)
$q=3/5\ K$ or $q=K$ CDW state, indicating the prediction of a CDW
state for TP VS$_2$ only exists within a narrow window of $U$ values.

\begin{figure}[tbp]
\includegraphics[width=\linewidth]{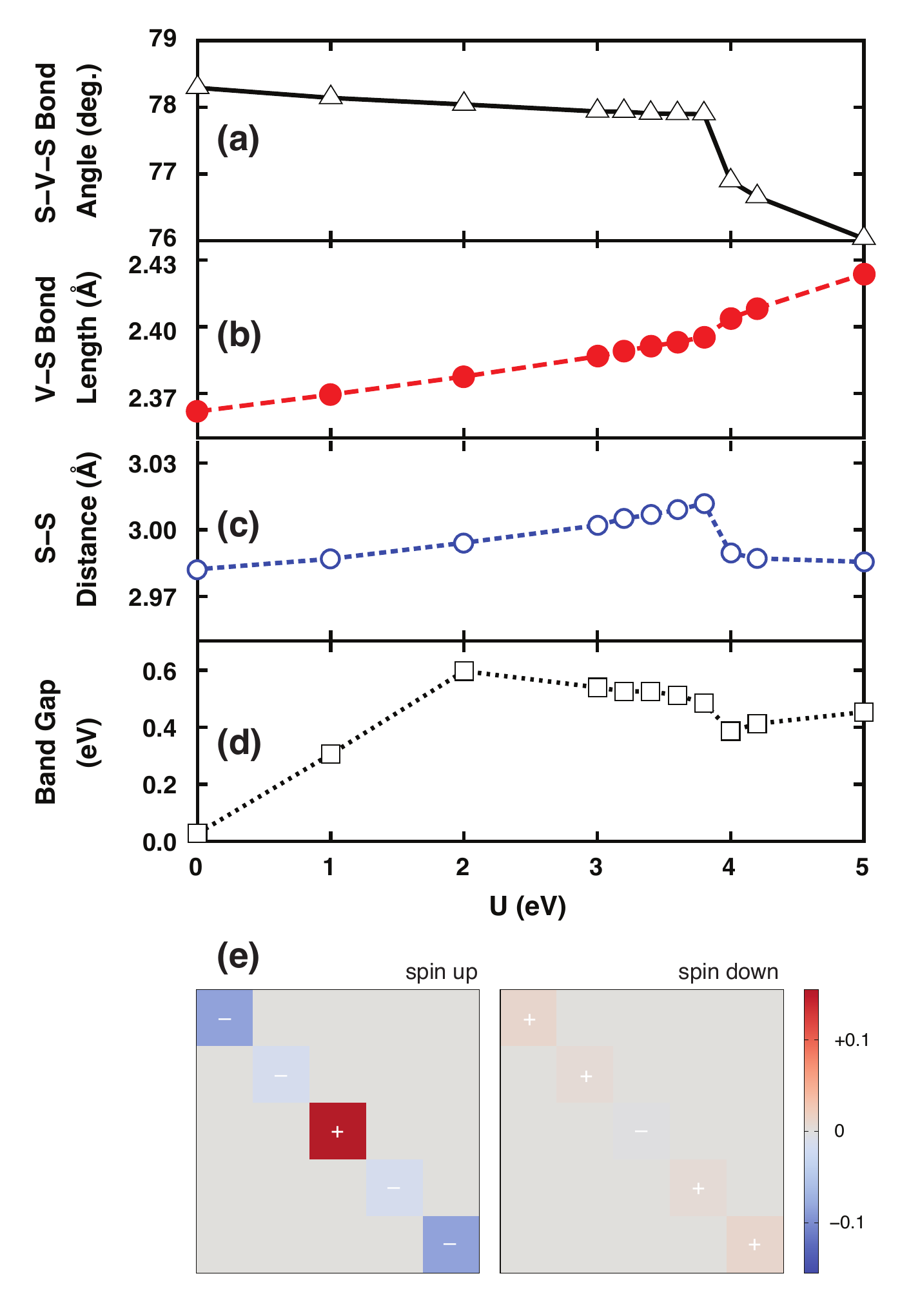}
\caption{(a) S--V--S bond angle, (b) V--S bond-length, (c)
  out-of-plane S--S distance, and (d) electronic band gap as a
  function of $U$ for FM TP VS$_2$. The density matrix difference for
  $U=4$ eV (ground state minus metastable state) for spin-up (left)
  and spin-down (right) electrons is displayed in panel (e). The
  matrix rows (columns) correspond to $d_{xy}$, $d_{yz}$
  $d_{3z^2-r^2}$, $d_{xz}$, and $d_{x^2-y^2}$ states from top to
  bottom (left to right).\label{u4_phase_trans}}
\end{figure}

For $U\ge4$ eV, both the $q=3/5\ K$ and $q=K$ soft modes disappear
(not pictured). Frozen phonon calculations indicate that the smallest
phonon frequency at $U=4$ eV is 126 cm$^{-1}$ for $q=K$, 97 cm$^{-1}$
for $q=3/5\ K$, and 79 cm$^{-1}$ for $q=1/2\ K$. In this regime of 4
eV $\le U<5$ eV we find that the $q=3/5\ K$ CDW phase is a separate
lower-energy state that exists in addition to the metastable
undistorted FM state.

The disappearance of the soft modes at $U\ge 4$ eV appears to be
related to a separate electronic and structural phase transition that
occurs within the primitive cell of FM TP VS$_2$. To describe the
phase transition, we plot in Fig. \ref{u4_phase_trans} several
structural parameters (out-of-plane S--V--S bond angle, V--S bond
length, and out-of-plane S--S distance) and the band gap as a function
of $U$ for FM TP VS$_2$. There is a sharp discontinuity in the
structural parameters at $U=4$ eV that most noticeably leads to
decreases in S--V--S bond angle and out-of-plane S--S distance. The
band gap shows a discontinuity and begins to decrease at $U=2$ eV when
the $A_1'$ level is no longer the lowest unoccupied state. At $U=4$ eV
there is a slight drop in band gap due to the phase transition, after
which it begins to increase roughly linearly. Using the relaxed
crystal structure from $U=4$ eV, we are able to converge a $U=4$ eV
DFT+$U$ calculation to a metastable state 6 meV higher in energy whose
electronic properties (e.g. density of states and local density
matrix) resemble those of lower $U$ (i.e., $U<4$ eV) as opposed to
this new ground state. This, along with the presence of
discontinuities in the structural and electronic properties, indicates
that the phase transition is of first order.

To better understand the electronic aspect of the phase transition, in
Fig. \ref{u4_phase_trans}(e) we plot the difference in the V on-site
density matrices (ground state minus metastable state) obtained using
the same crystal structure. The most significant changes occur in the
spin-up channel. Compared to the metastable state, in this spin
channel the ground state has 0.16 additional occupancy of the $A_1'$
($d_{3z^2-r^2}$) state and 0.16 less in total occupancy of the $E'$
($d_{x^2-y^2}$ and $d_{xy}$) states.

Given the crude nature of DFT+$U$, one must view these results with
caution. More advanced calculations using DFT+DMFT, in addition to
experiments, would be needed to judge the veracity of this predicted
CDW. A smaller value of $U$ might be more relevant in VS$_2$ to
compensate for errors associated with Hartree-Fock treatment of the
impurity problem.


\subsection{DFT+$U$ relative phase stability}

\begin{figure}[tbp]
\includegraphics[width=\linewidth]{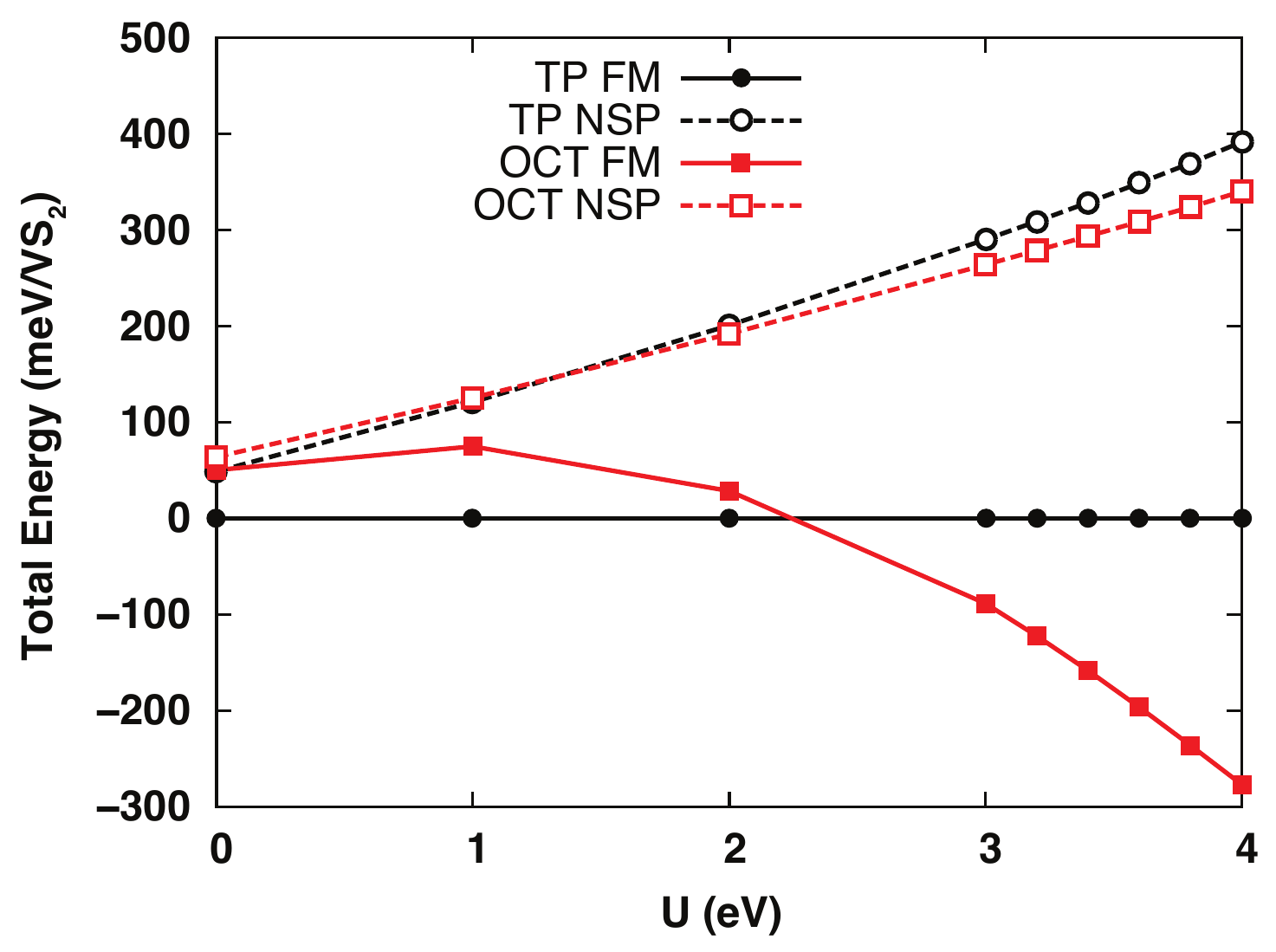}
\caption{Total energy of NSP TP (black dashed line and open circles),
  NSP OCT (red dashed line and open squares), and FM OCT (red solid
  line and filled squares) states referenced to the FM TP (black solid
  line and filled circles) state energy as a function of
  $U$.\label{dftu_energy_levels}}
\end{figure}


To explore the impact of $U$ on the relative energetics of the TP and
OCT phases, in Fig. \ref{dftu_energy_levels} we show the total energy
of the NSP and FM states for TP and OCT VS$_2$ referenced to the TP FM
state energy. Here the CDWs are not considered as they do not change
the qualitative results; all calculations are in the primitive
cells. For $U=0$ the TP FM state is the ground state with the TP NSP,
OCT FM, and OCT NSP states 49, 50, and 64 meV higher in energy,
respectively. As $U$ increases the NSP states are each monotonically
destabilized by several hundreds of meV compared to the TP FM state as
expected. The OCT FM phase has a more complicated nonmonotonic
behavior, initially slightly increasing its relative energy with $U$
and then decreasing its relative energy for $U>1$ eV. For $U$ values
larger than 1 eV the OCT FM state becomes an insulator with the
$A_{1g}$ state fully polarized (V magnetic moment of 1 $\mu_B$) and is
energetically stabilized; for $U=3$ eV it is lower in energy than the
TP FM state by 88 meV, and the energy stabilization increases upon
further increasing $U$.





\begin{figure}[tbp]
\includegraphics[width=\linewidth]{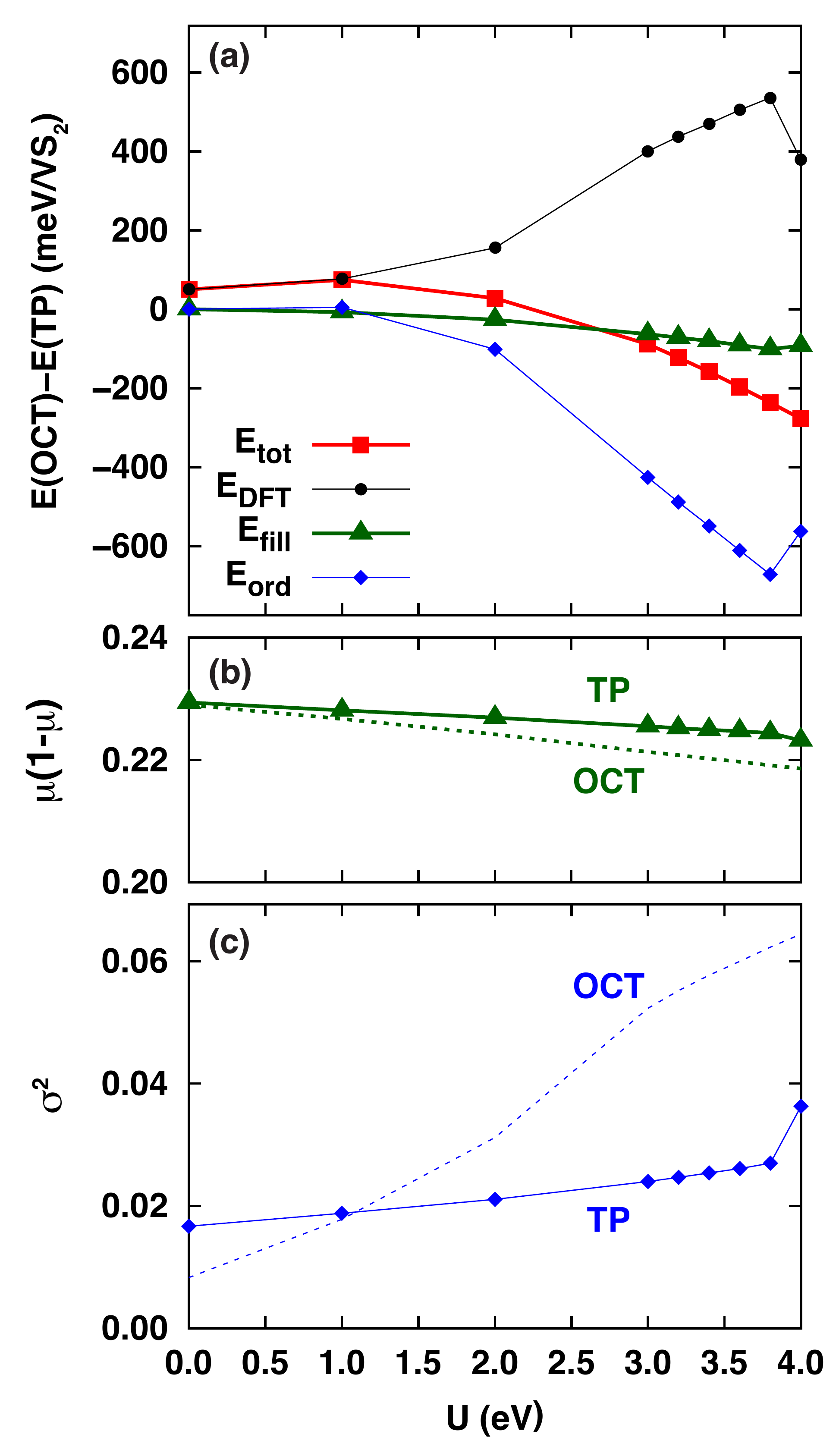}
\caption{(a) DFT+$U$ total energy of FM OCT phase minus that of FM TP
  phase (red squares) and decomposition into DFT (black circles),
  filling (green triangles), and ordering (blue diamonds)
  contributions as a function of $U$. (b) $\mu(1-\mu)$ (green) and (c)
  $\sigma^2$ (blue) as a function of $U$. Solid (dashed) lines with
  (without) symbols correspond to the TP (OCT) phase in panels (b) and
  (c).
\label{dftu_filling_ordering}}
\end{figure}

To gain further insight into the stabilization of FM OCT over FM TP
VS$_2$ with $U$, we introduce a new spectral decomposition of the
DFT+$U$ energy functional into contributions from DFT ($E_{DFT}$),
filling of V $d$ orbitals ($E_{fill}$), and ordering of V $d$ orbitals
($E_{ord}$):
\begin{align*}
&E_{DFT+U}=E_{DFT}+E_{fill}+E_{ord} &&
\\
&E_{fill}=U(2l+1)\mu(1-\mu) &&
E_{ord}=-U(2l+1)\sigma^2
\end{align*}
where $l$ is the angular momentum ($l=2$ for $d$ electrons) and $\mu$
and $\sigma$ are the mean and standard deviation of the eigenvalues of
the local $d$ density matrix. The filling and ordering terms added
together give the standard interaction and double counting terms in
DFT+$U$ for $J$ set to 0. This decomposition provides a convenient way
to isolate and quantify the contributions of the average filling of
the $d$ shell and the spin and orbital ordering of the $d$ shell to
the interaction and double counting energetics. The former elucidates
the energetics associated with moving charge into or out of the
correlated subspace, while the latter is the means by which
Hartree-Fock captures the energetics of electronic correlations.



As shown in Fig. \ref{dftu_filling_ordering}(a), for $U=1$ eV
$E_{DFT}$ (black circles) and $E_{ord}$ (blue diamonds) are
responsible for the further stabilization of the TP phase compared to
$U=0$. For larger $U$, the $E_{fill}$ term (green triangles)
increasingly favors the OCT phase by as much as 101 meV as $U$
increases. The total E(OCT)--E(TP) (red squares) decreases with $U$ a
factor of 3 to 4 faster than $E_{fill}$. $E_{DFT}$ and $E_{ord}$ tend
to oppose each other, but overall the negative $E_{ord}$ term is
dominant and this contributes significantly to the overall
stabilization of the OCT phase. The $E_{ord}$ and $E_{DFT}$ terms
increase in magnitude significantly faster once the OCT phase becomes
an insulator at $U=2$ eV. We find the same qualitative behavior when
we freeze the ions at the $U=0$ structures, indicating this is not an
effect of structural relaxation.

The filling factor $\mu(1-\mu)$ and the ordering factor $\sigma^2$ are
plotted for both phases in Fig. \ref{dftu_filling_ordering}(b) and
Fig. \ref{dftu_filling_ordering}(c), respectively. Interestingly, the
TP and OCT phases have an almost identical filling of the V $d$ shell
with $\mu(1-\mu)=0.229$ at $U=0$. On the other hand, the $\sigma^2$
terms are substantially different at $U=0$: $\sigma^2$ is 0.0167 in
the TP phase as opposed to only 0.0083 in the OCT phase. This stems
from the complete spin polarization of the $A_1'$ state in the TP
phase, as opposed to the partial spin polarization in the OCT
phase. The preceding statement can be supported by investigating the
NSP state for both the TP and OCT phases for $U=0$, which yields much
more similar $\sigma^2$ values of 0.0037 and 0.0047,
respectively. Therefore, the pure crystal fields in each respective
case results in a similar and small $\sigma^2$, while the differing
degrees of spin polarization are responsible for the large initial
difference at $U=0$. This enhanced spin ordering in the TP phase leads
to the enhanced stabilization of the TP phase in the limit of small
$U$ since $\partial E_{ord}/\partial U\sim-\sigma^2$ and because the
initial fillings are nearly identical. However, this trend is only
guaranteed for small $U$ and as we pointed out above the trend
reverses for $U>1$ eV. We therefore proceed to examine each
contribution as a function of $U$. In terms of the filling
contribution, the OCT phase filling factor decreases with $U$ twice as
fast as it does for the TP phase for $U\le3.8$ eV. The $\sigma^2$ for
the OCT phase increases 5.2 times as fast as does that of the TP phase
for $U\le3.8$ eV, since both the $A_{1g}$ and the $E_g'$ states are
polarizable, and for $U=3.8$ eV it has an ordering factor 2.3 times as
large. Therefore, both the decreased filling and increased ordering of
the $d$ orbitals of the OCT phase contribute to its stabilization for
larger $U$.

\subsection{Possibility of realizing TP VS$_2$}

Only the OCT phase of VS$_2$ has been observed experimentally, in bulk
and nanosheet
forms.\cite{murphy_preparation_1977,feng_metallic_2011,feng_giant_2012,zhong_ferromagnetism_2014,gauzzi_possible_2014}
DFT predicts the TP phase is the thermodynamic ground state, while
DFT+$U$ predicts that the OCT phase becomes the ground state when $U$
surpasses a moderate value of approximately $2.3$ eV. More advanced
calculations, including DFT+DMFT and possibly cluster extensions of
DMFT, will be needed to definitively settle this issue from a
theoretical standpoint. Given that TP may in fact be the ground state,
or possibly a metastable state sufficiently low in energy to be
achieved experimentally, we explore possible reasons why it has not
been observed in experiment.

The initial synthetic route to VS$_2$ was delithiation from
LiVS$_2$.\cite{murphy_preparation_1977} This lithiated compound has a
layered octahedral structure.\cite{vanlaar_preparation_1971}
Therefore, one possibility is that VS$_2$ is stuck in a metastable OCT
state. Within DFT, we compute an energy barrier of 0.69 eV per formula
unit based on a linear interpolation between the TP and OCT monolayer
structures allowing only out-of-plane ionic relaxation. This value is
in agreement with nudged elastic band calculations that found a
barrier of 0.66 eV.\cite{zhang_dimension-dependent_2013} The large
barrier supports the possibility that it could very challenging to
change phases. Another high-temperature synthesis technique did not
use LiVS$_2$ but still resulted in the OCT
phase.\cite{ohno_x-ray-absorption_1982,ohno_x-ray_1983} One
possibility is that finite temperature plays a role in destabilizing
the TP phase since there is evidence that the phonon entropy is
greater for the OCT phase.\cite{zhang_dimension-dependent_2013}

A more recent high-pressure synthesis of VS$_2$ also yielded the OCT
phase.\cite{gauzzi_possible_2014} We performed spin-polarized DFT
(i.e., $U=0$) calculations of bulk VS$_2$ under pressure and find that
for sufficiently high pressure the OCT phase becomes the ground state,
so this could be responsible for why the TP phase is not observed. In
these calculations we considered 2H$_c$ (MoS$_2$-like)
stacking\cite{katzke_phase_2004} for the TP phase and O1
(CoO$_2$-like) and O3 (LiCoO$_2$-like)
stackings\cite{van_der_ven_first-principles_1998} for the OCT
phase. At 5 GPa the TP phase is still the ground state but only 15 meV
lower in energy compared to 50 meV for 0 GPa. At 10 GPa the TP phase
becomes 26 meV higher in energy than the OCT phase. Based on these
observations, if the TP phase is the ground state we predict that
synthesis under ambient pressure, low temperature, and not involving a
LiVS$_2$ precursor will be most effective to attempt to realize TP
VS$_2$.

\section{Conclusions}

We have demonstrated that monolayer TP VS$_2$ has an isolated
low-energy band at level of NSP DFT, which arises due to a combination
of the TP crystal field and the nearest-neighbor V--V
hopping. Including spin polarization reveals that the exchange is
ferromagnetic and yields a FM insulator with a small band gap. Other
spin configurations result in metallic states substantially higher in
energy, indicating that spin-dependent DFT is not putting VS$_2$ in
the Mott regime. While TP VS$_2$ has not been observed in experiment
in any form, spin-polarized DFT does predict it is lower in energy
than the OCT phase. DFT captures the known CDW in the OCT phase, which
strongly diminishes the magnetism relative to the undistorted phase.
However, DFT appears to grossly overestimate the CDW amplitude in this
phase. Specifically, the V--V bond length differences from DFT are far
larger than those of the existing XAFS study.\cite{sun_in-situ_2015}

Accounting for local correlations via DFT+$U$ produces a $S=1/2$ FM
insulating state in the TP phase, which is in the Mott regime for
moderate values of $U$. For a small regime of finite $U$, we find a
CDW in the TP phase at $q=3/5\ K$. For the OCT phase, increasing $U$
diminishes the amplitude of the CDW. For the ferromagnetic CDW state,
the amplitude decreases slowly before rapidly collapsing near $U=3$
eV. However, for this regime of $U$, magnetism with anti-aligned spins
becomes energetically favored over ferromagnetism. In this magnetic
configuration we find metallic behavior as in experiments and the V--V
bond length differences of the CDW phase are within reasonable
comparison to XAFS experiments.

Regarding relative phase stability, above a reasonably small $U$
(approx. 2.3 eV) the energy ordering of TP and OCT phases reverses
with the OCT phase becoming the ground state. More advanced
calculations, including DFT+DMFT and possibly cluster extensions of
DMFT, will be needed to settle which is the ground state structure and
determine whether the CDW in the TP phase is physical.

If the TP phase can be realized, it has the potential for novel
physics: it would be a rare example of a $S=1/2$ Mott insulator on a
triangular lattice with strong FM correlations. Its monolayer nature
might enable doping via gating, allowing one to probe the doped Mott
insulator in a precise fashion without simultaneously introducing
disorder.




\begin{acknowledgments}
This research used resources of the National Energy Research
Scientific Computing Center, a DOE Office of Science User Facility
supported by the Office of Science of the U.S. Department of Energy
under Contract No. DE-AC02-05CH11231. The authors acknowledge support
from the NSF MRSEC program through Columbia in the Center for
Precision Assembly of Superstratic and Superatomic Solids
(DMR-1420634). E.B.I. gratefully acknowledges support from the
U.S. Department of Energy Computational Science Graduate Fellowship
(Grant No. DE-FG02-97ER25308).
\end{acknowledgments}

\bibliography{main}

\end{document}